\documentclass[letterpaper,aip, amsmath,amssymb,reprint]{revtex4-1}

\usepackage{graphicx}% Include figure files
\usepackage{dcolumn}% Align table columns on decimal point
\usepackage{bm}% bold math
\usepackage[utf8]{inputenc}
\usepackage[T1]{fontenc}
\usepackage{mathptmx}
\usepackage{etoolbox}
\usepackage{caption}
\usepackage{xcolor}
\usepackage{fancyhdr}
\usepackage{etoolbox}

\usepackage[flushleft]{threeparttable}
\usepackage[title]{appendix}

\graphicspath{ {./Figures/}}

\renewcommand{\arraystretch}{1.25}
\makeatletter
\def\@email#1#2{%
 \endgroup
 \patchcmd{\titleblock@produce}
  {\frontmatter@RRAPformat}
  {\frontmatter@RRAPformat{\produce@RRAP{*#1\href{mailto:#2}{#2}}}\frontmatter@RRAPformat}
  {}{}
}

\begin{document}
\fancyfoot[L]{\textit{Preprint submitted to arXiv}}
\preprint{AIP/123-QED}

\title{Supersonic Microparticle Impact Experiments at Temperatures Approaching 2000 °C}
% Force line breaks with \\
\author{Jamshid Ochilov}
\author{Isaac Faith Nahmad}
\author{Intekhab Alam}
\author{Peter Yip}
\author{Suraj Ravindran}
\email[Author to whom correspondence should be addressed: ]{sravi@umn.edu}
\affiliation{Department of Aerospace Engineering and Mechanics, University of Minnesota, Minneapolis, MN 55455}


\begin{abstract}
Experiments at extreme strain rates and temperatures are critical for characterizing materials in high-speed applications. In this study, we develop a laser-driven particle impact platform capable of accelerating microparticles to supersonic velocities and impacting targets heated to temperatures approaching 2000 °C. The conventional laser-induced particle impact testing (LIPIT) system has been modified to enable high-temperature experiments through the integration of a resistive heating system and the development of a robust launch pad assembly suitable for accelerating particles in high-temperature environments. To eliminate the oxidation of materials at elevated temperatures, an optically accessible portable vacuum chamber has been developed and integrated into the setup. The capabilities of the system are demonstrated through a study of the temperature dependent particle impact cratering behavior of POCO graphite. With this new platform, high-velocity, high-temperature impact experiments can be performed in a controlled environment, supporting the investigation of materials under extreme conditions.
\end{abstract}
\maketitle
\section{Introduction}
Materials employed in high-speed applications are subjected to extreme (and often simultaneous) mechanical and thermal loads, characterized by high-strain-rate deformations and elevated temperatures. High-speed particle impact exemplifies these extreme conditions and is encountered in various critical engineering systems. For instance, in hypersonic flight, aircraft surfaces under extreme aerothermal loads can experience particle impact erosion and spallation damage at high temperatures when subjected to particle-laden flow  \cite{connolly2021simulations, habeck2024characterization}. Comparable damage mechanisms are observed in gas turbine and rocket engines, where foreign object damage (FOD) occurs in high-temperature environments \cite{neilson1968experimental, chen2003foreign}. It is well established that materials generally exhibit strain-rate hardening and thermal softening; however, the concurrent influence of these effects can significantly alter the governing mechanisms of deformation and failure \cite{grunschel2007dynamic, zaretsky2012impact, dowding2024metals}. Therefore, a thorough understanding of material behavior under combined high-strain-rate and high-temperature loading is critical for developing material models to predict the deformation and damage behavior in extreme environments.

Experimental techniques have been developed and implemented over the years to study the high-temperature and high-strain-rate response of materials. Among these, split-Hopkinson pressure bars (SHPB), or Kolsky bars, equipped with high-temperature heating systems, are widely used in characterizing material behavior at strain rates ranging from 10$^2$ to 10$^4$ /s, and temperatures up to 1600 °C \cite{pittman2024method,lew2022joule, kidane2008dynamic,nemat1997direct,zhang2020study}. In higher strain rate regimes (above 10$^4$ /s), high-temperature plate impact experiments are used to probe the dynamic strength and equation of state (EOS) of materials \cite{frutschy1998high,zaretsky2012effect, zuanetti2018mechanical, wang2017shock}. Gas-driven plate impact and SHPB experiments offer robust platforms for rigorous exploration of material response under extreme conditions and provide high-fidelity data on the rate, pressure, and temperature-dependent deformation and damage behavior of materials \cite{ravindran2021dynamic, gandhi2022dynamic, kettenbeil2020pressure}. However, their capacity for high-throughput experiments remains limited.

In particle impact erosion and FOD encountered in high-speed applications, materials are subjected to strain rates exceeding $10^5$ /s and temperatures above 1000 °C. Bulk erosion of materials at elevated temperatures has been studied using centrifugal accelerators and sand blast-type rigs for low velocity \cite{tabakoff1979test, levy1986elevated, chinnadurai1995high, zhou1995erosion, lindsley1995design, kulu2005solid}, and hypersonic wind tunnels (with particulates injected into the flow) as well as gas-stream accelerators for high-velocity particle impact \cite{wakeman1979erosion, lorenz1970simulation}. Although bulk erosion experiments provide valuable macroscale weight loss and surface recession rates, they are inadequate for investigating the microscale mechanisms in individual particle impact damage and cratering.

Traditionally, ballistic and hypervelocity gas gun facilities have been used to study material damage and cratering behavior under single projectile impact \cite{choi2014foreign, rubin1974influencer, adler11979erosion, hebert2017hypervelocity}. However, gas launch systems typically accelerate millimeter-sized particles, while many erosion and FOD phenomena involve particles on the order of microns. In addition, elucidating the full extent of erosion mechanisms and the development of predictive models requires extensive experimental data involving large test matrices, which are challenging to accomplish and obtain from conventional experiments given their high cost and low-throughput nature.

An alternative experimental technique in probing materials at extreme strain rates is the laser-induced particle impact test (LIPIT), developed over a decade ago \cite{lee2012high}. The technique operates via a driving mechanism similar to that of laser-induced forward transfer (LIFT) \cite{bohandy1986metal, schultze1991laser} and laser-driven flyer experiments \cite{paisley1988laser, trott1989acceleration, brown2012simplified}. Using LIPIT, individual or a cluster of microparticles ranging in size from 1 µm to 500 µm can be accelerated to high velocities (up to 4.2 km/s for 3.7 µm particles \cite{veysset2021high}) using a high-energy pulsed laser \cite{lee2014dynamic, thevamaran2016dynamic, hassani2018situ, veysset2020laser, cai2020superior, veysset2021high, portela2021supersonic, dhiman2021microscale, lee2023studies, kajihara2024development, tang2024strength, taylor2025particle, song2025dynamic, ochilov2025laser}. These experiments are equipped with high-magnification, high-speed imaging systems to measure the impact and rebound velocities of the projectiles and to observe the evolution of impact damage and spallation/ejecta characteristics. The resulting crater morphology and dimensions are characterized and measured postmortem using 3D profilometry. They can be used together with impact and rebound velocities to estimate the hardness and yield strength of materials \cite{sun2020transition, hassani2020material}. A key advantage of this technique is its high-throughput capability, allowing for the rapid completion of numerous experiments under varying impact conditions while generating particle-impact erosion-type loading with relative ease.
Recently, LIPIT experiments have been performed at elevated temperatures up to
300 °C \cite{reiser2023microparticle}, achieved through the implementation of launch pad designs with alternative expansion/driving layers to traditional elastomers.  Efforts have also been made to implement launch pad designs with configurations and driving mechanisms for improved performance in particle size and launch velocity, spatial precision consistency, and limiting debris \cite{veysset2020laser, reiser2023microparticle, chen2025high, portillo2025high}.
In this study, a LIPIT system is developed that is capable of particle impact experiments at ultrahigh temperatures in atmosphere and under vacuum. Target heating is achieved via direct resistive (Joule) heating, and metallic foils (of aluminum and copper) are implemented as launch pad expansion/driving layers for high-temperature experiments. The system capabilities are demonstrated through a case study on ultra-fine grain POCO graphite.

\begin{figure}[t]
\includegraphics[width=0.45\textwidth]{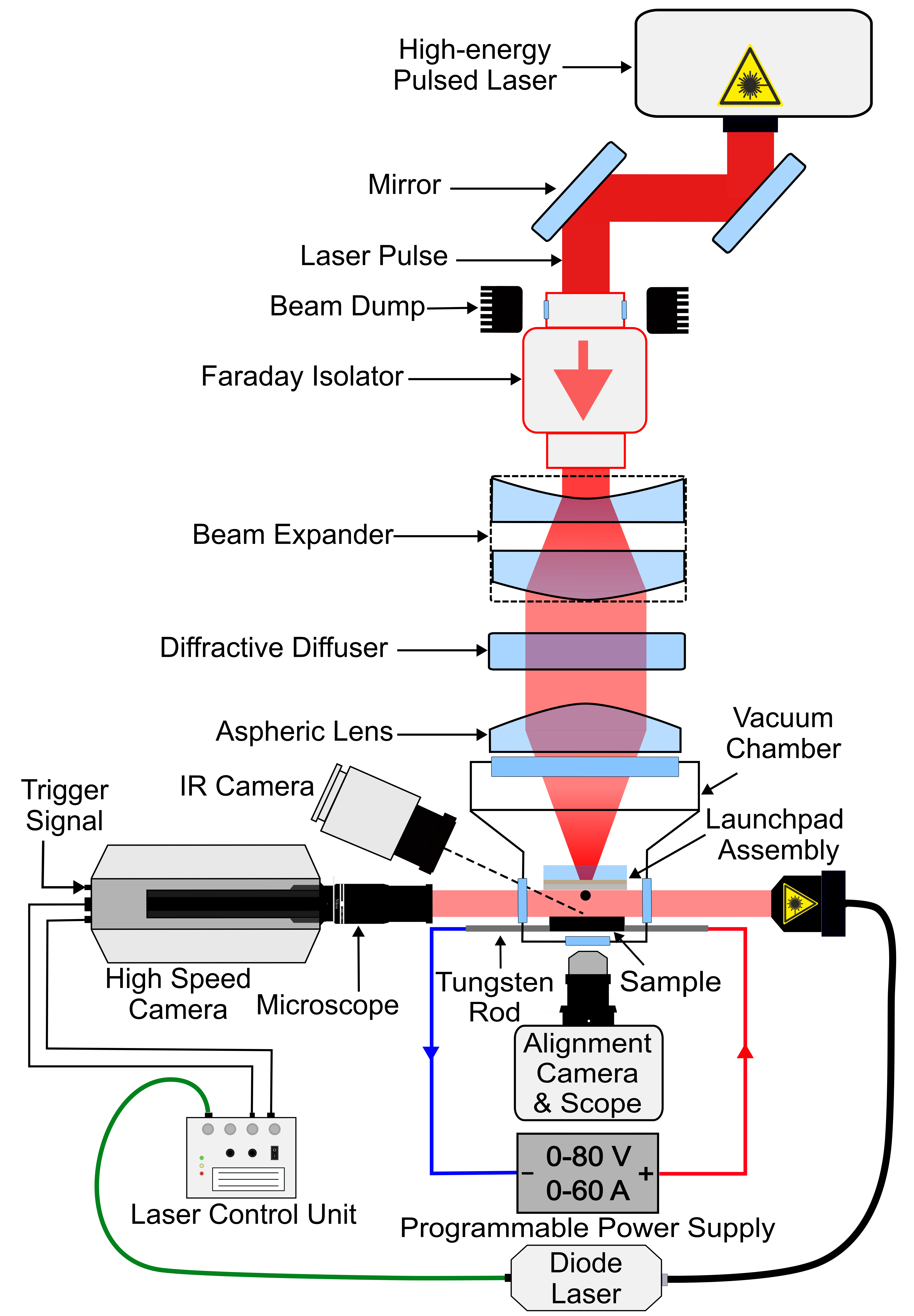}
	\caption{Schematic of the experiment setup.}
	\label{fig:ExpSetupSchem}
\end{figure}

\section{Materials and Methods}\label{ExperimentalMethods}
\subsection{Experimental Methods}

The experiment setup (schematic shown in Fig. \ref{fig:ExpSetupSchem}) is an optical tabletop system comprising several subsystems: a high-energy Q-switched laser and beam manipulation optical train, launch pad and target assembly fixtures, vacuum chamber assembly, illumination and imaging systems for particle alignment and high-speed diagnostics. To enable high temperature experiments, a resistive heating system is incorporated in the setup along with a methodology to accelerate particles in high-temperature experiments. The driving laser implemented here is the Amplitude Powerlite DLS Plus. It is a Q-Switched Nd:YAG laser capable of releasing individual 4 to 9 ns pulses with energies up to 3 J at 1064 nm wavelength and 12 mm beam diameter.

\begin{figure*}
\centering
\includegraphics[width=0.91\textwidth]{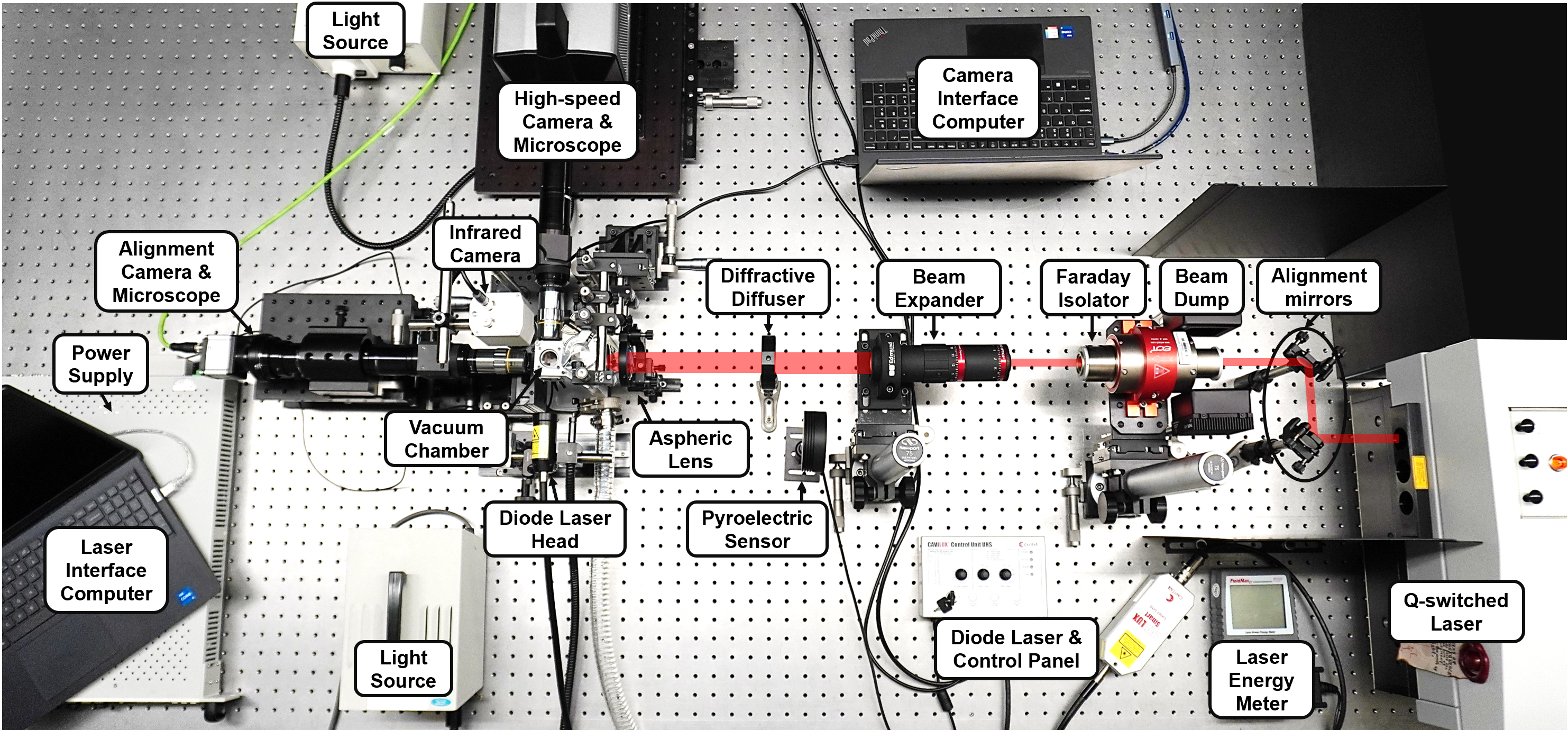}
	\caption{Particle impact (LIPIT) facility with experiment capabilities at ultra-high temperatures and under high vacuum.}
	\label{fig:ExpSetup}
\end{figure*}

\subsubsection{Beam Manipulation Optical Train}
The 12 mm laser beam released from the main laser is aligned to a predetermined experiment axis using a Z-fold configuration of a pair of 25.4 mm (1") diameter mirrors (Edmund Optics $\#38-909$) using kinematic mounts (see Fig. \ref{fig:ExpSetup}). Next, the beam is passed through a Faraday isolator (EOTech PAVOS) to block any back reflections from reaching the laser. The Faraday isolator rotates the polarization of the laser beam by 45 degrees and requires the use of safety beam dumps at the input polarized escape ports when the laser is operated at a pulse energy of 1 J or higher. The beam diameter is then increased from 12 mm to 30 mm using a variable beam expander (Edmund Optics $\#87-568$). This is done to reduce the fluence at the optical components down the range, thereby avoiding damage and optimizing the size and quality of the focused beam spot at the launch pad. It is noted that a Galilean design of expander is chosen to prevent the beam from converging to a point where it would risk ionizing the air within the expander enclosure at higher-energy pulses. Following the expansion, the Gaussian beam is homogenized into a circular near-flat-top (top-hat) distribution using a diffractive diffuser (Holo/Or RD-294-I-Y). This spatially homogenized distribution of the beam has previously been implemented in laser-driven flyer experiments \cite{brown2012simplified, mallick2019laser}, and is chosen here because it was found to be preferable to Gaussian beams when used with the current launchpad configuration. At this point along the beam path, a neutral density (ND) filter can optionally be introduced to reduce the pulse energy to a few millijoules. This would enable the use of commonly employed (polymer-based expansion layer) designs for launch pad assemblies in room-temperature experiments. Finally, the beam is focused to a 500 - 700 µm diameter (varying depending on the driving layer material and thickness) spot on the glass substrate-ablative layer interface of the launch pad using a 100 mm focal length, 50.8 mm (2") diameter aspheric lens (Thorlabs AL50100J-C). The larger spot sizes are required to produce sufficient ablation products to generate confined pressures enough to rapidly deform/expand the metal foils.  The lens is mounted on a 6-axis locking kinematic mount (Thorlabs K6X2) to enable precise alignment of the lens center with the beam and to achieve a circular focused spot using the tilt and tip adjustments. The laser pulse energy is measured for each experiment using a pyroelectric sensor (Coherent FieldmaxII-TOP).

\subsubsection{Launch Pad Assembly: Design for Extreme Temperatures}
Traditional launch pad assembly designs for LIPIT have predominantly utilized an elastomeric expansion layer, such as polydimethylsiloxane (PDMS) or polyurea (PU), spin-coated over a thin metallic ablation layer deposited on a borosilicate glass substrate \cite{lee2014dynamic, thevamaran2016dynamic, hassani2018situ, veysset2020laser, cai2020superior, portela2021supersonic, dhiman2021microscale, lee2023studies}. However, the relatively low degradation temperature of elastomers limits their application as expansion layers to room temperature experiments \cite{reiser2023microparticle, camino2001polydimethylsiloxane, temizkan2021preparation}.

\begin{figure}[hbt!]
\includegraphics[width=0.46\textwidth]{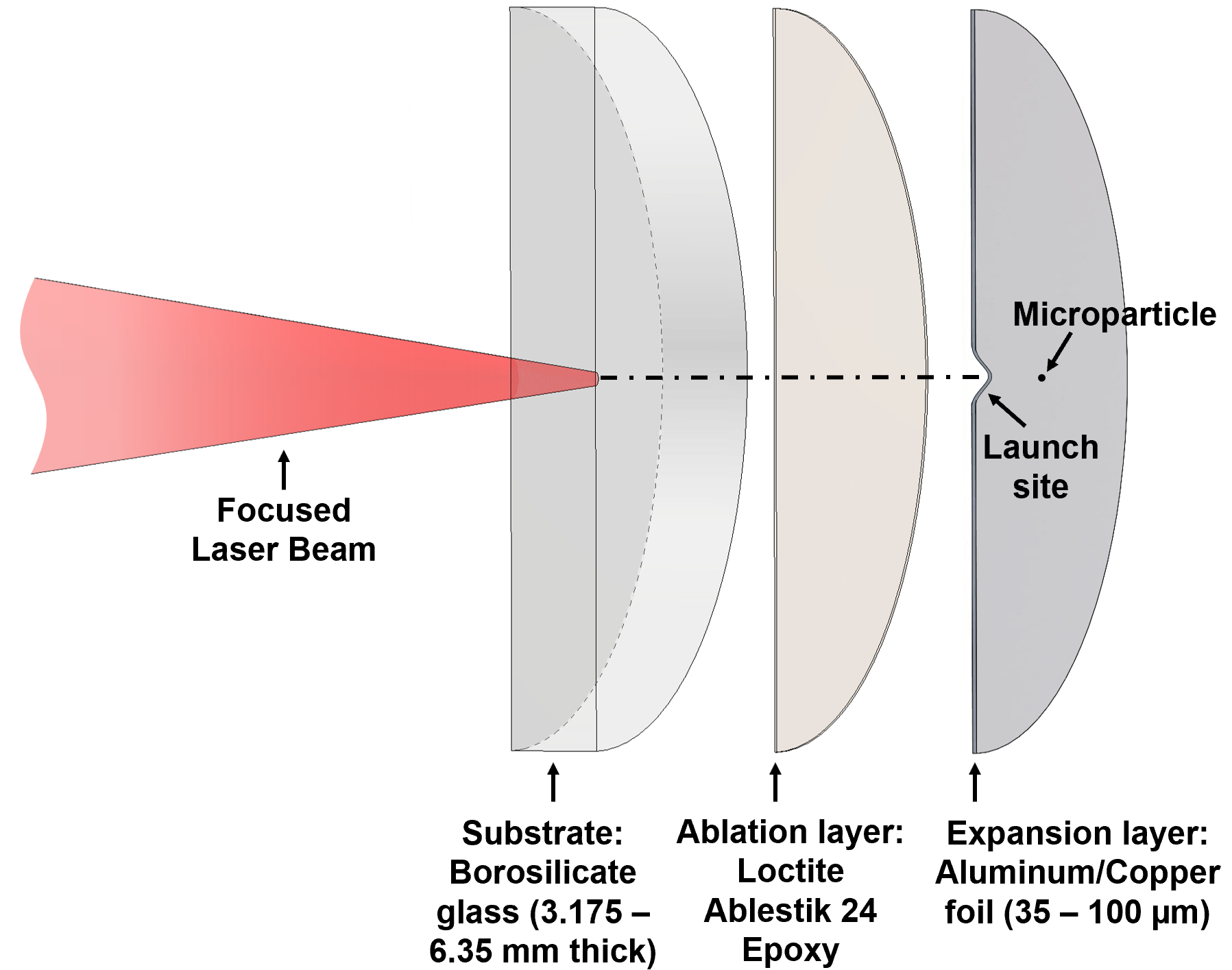}
	\caption{Launch pad assembly: exploded view of the layup sequence cross-section shown along with the incident beam and launch mechanism.}
	\label{fig:layup}
\end{figure}

This is due to the proximity of the target sample to the surface of the launch pad, typically positioned within a submillimeter spacing, where high-temperature experiments directly lead to heating of the expansion layer.

\begin{figure}[b]
\includegraphics[width=0.49\textwidth]{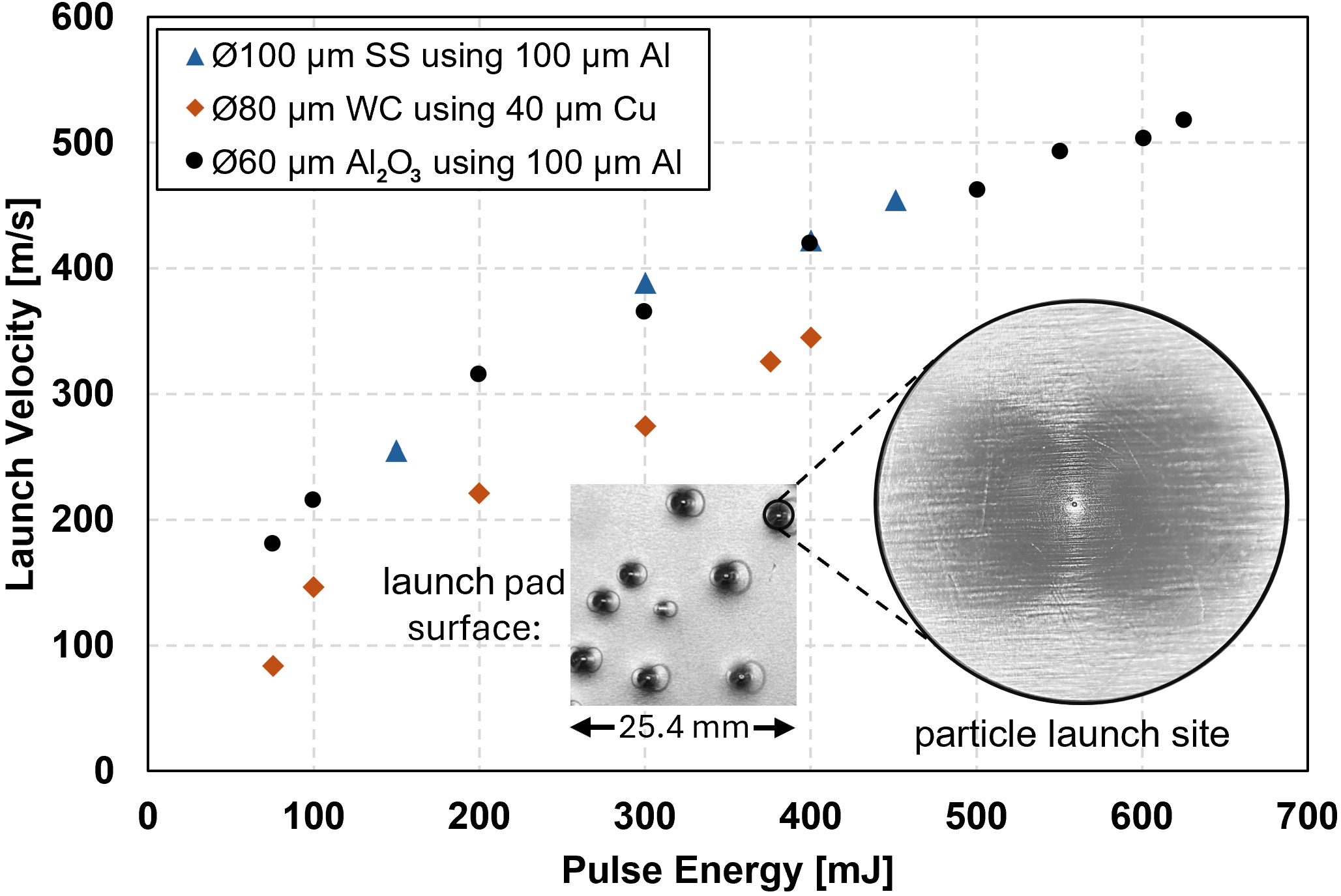}
	\caption{Particle launch velocity vs. laser pulse energy for 100 µm aluminum and 40 µm copper foil as expansion layers driving nominally 60 µm alumina, 80 µm tungsten carbide, and 100 µm stainless steel  microspheres. 25.4 $\times$ 25.4 mm$^2$ launch pad surface containing 8 post-mortem launch sites, and a magnified launch site is overlaid.}
	\label{fig:VvsE}
\end{figure}

\begin{figure*}
\includegraphics[width=0.85\textwidth]{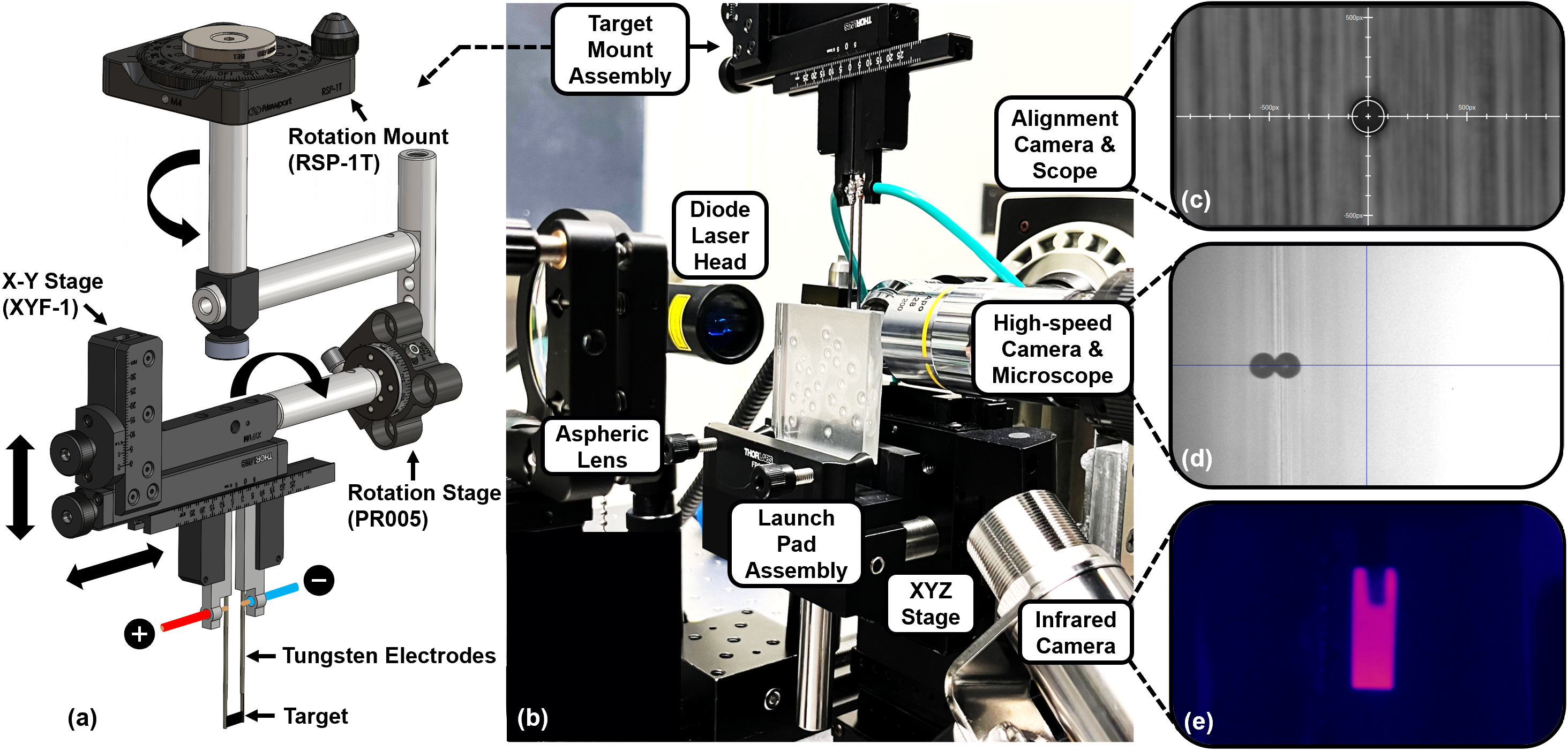}
	\caption{(a) Target mount assembly for experiments in the atmosphere. (b) Launch pad and target mount experiment configuration. (c) Rear particle alignment camera and (d) high-speed camera view of a 60 µm particle on launch pad surface. (e) IR camera view of the mounted target sample.}
	\label{fig:targetMount}
\end{figure*}

Glass-ablative-glass configurations previously developed and implemented were demonstrated to achieve high spatial precision and experiments at target temperatures of up to 300 °C \cite{reiser2023microparticle}. The limitations were reported to be the fracturing of the front layer of glass due to the deterioration of the adhesive at elevated temperatures along with the fragmenting of ceramic particles and the plastic deformation of metallic particles at launch due to the extreme accelerations exerted by the stiff glass layer. In the designs proposed in this study, these issues are overcome by using ductile metal foils that enable the driving of particles to supersonic speeds at extreme temperatures without fragmenting/deforming launched particles.

A 100 µm thick aluminum foil (Alufoil Inc.) is adhesively bonded to a 6.35 mm (1/4") thick borosilicate glass substrate (McMaster-Carr) using a thin layer of epoxy adhesive (Henkel Loctite Ablestik 24) that also acts as the ablation layer. An exploded view of the launch pad assembly cross-section is shown in Fig. \ref{fig:layup}, where the expansion launch mechanism of the metallic layer is illustrated. Microparticles are placed on the free surface of the metallic foil. This glass-epoxy-metal foil design has been found to enable experiments at extreme temperatures while also being easier to fabricate than traditional designs, which require processes such as metallic layer deposition (CHA e-beam deposition or sputtering) and the spin coating of the elastomeric layer. A 50 µm thick copper foil has been found to perform comparably as an expansion layer as an alternative to 100 µm aluminum, with the added advantage of higher thermal stability. The use of metallic expansion layers also enables the launch of microparticles of larger sizes and highly dense materials, such as tungsten carbide, which are challenging to accelerate to supersonic speeds when using elastomeric expansion layers. 
Plotted in Fig. \ref{fig:VvsE} are the launch velocity vs. incident pulse energy data for aluminum (100 µm) and copper (40 µm) expansion layers driving nominally 60 µm alumina, 80 µm tungsten carbide, and 100 µm stainless steel microspheres without causing fragmentation or plastic deformation. It is noted that the higher pulse energies are required for the ablation of the epoxy layer to produce sufficient confining pressure to rapidly deform and expand the metallic foils.

\begin{figure}
\includegraphics[width=0.48\textwidth]{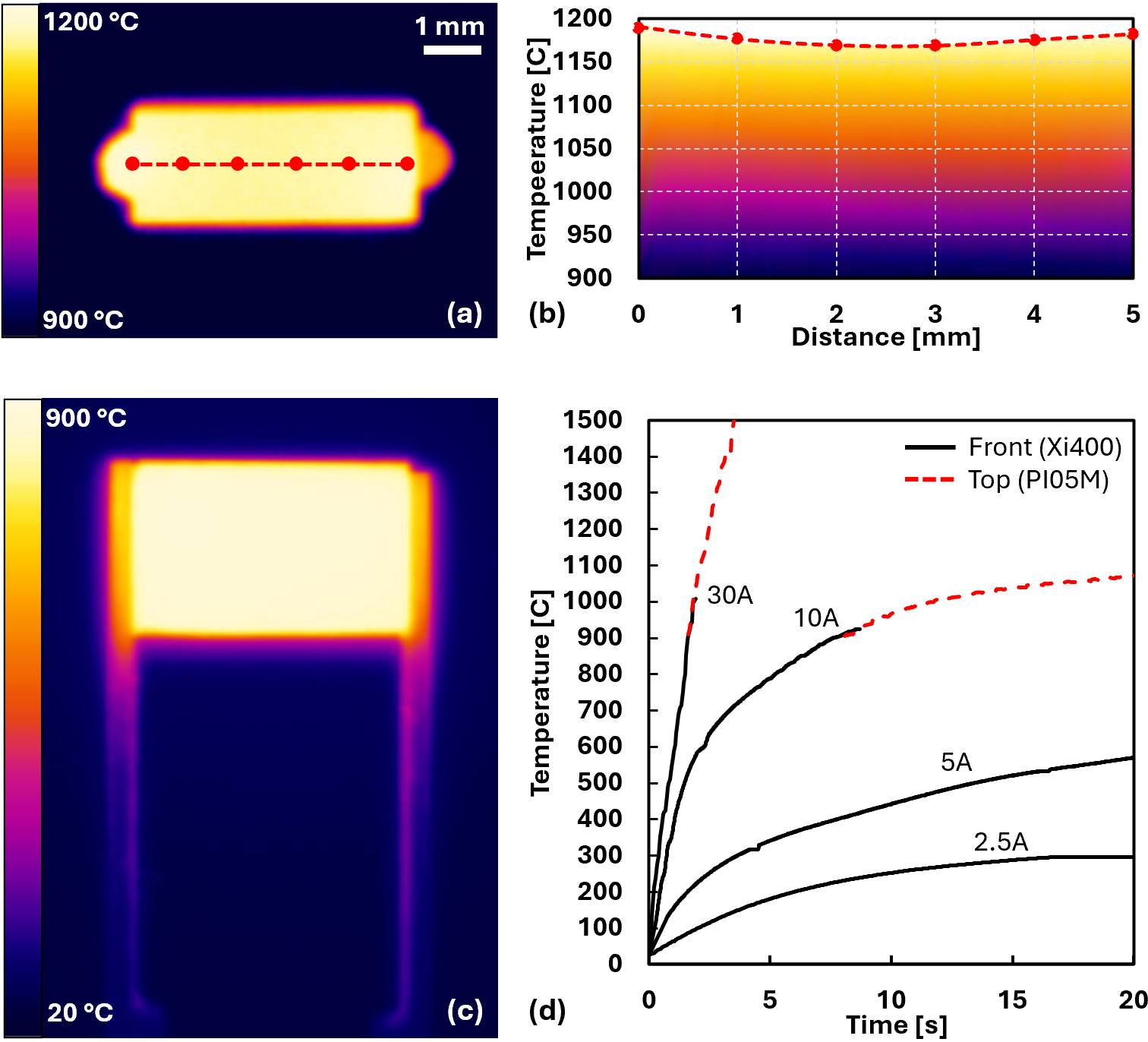}
	\caption{Graphite (POCO ZXF-5Q) material resistive heating calibration. (a) Top view (Optris PI05M) and (c) front view (Optris Xi400). (b) Top surface temperature variation along the centerline plotted on top of the gradient extracted from (a). (d) Temperature vs. time plots for varying current loads - measured at the top and front surface center points.)}
	\label{fig:tempCalib}
\end{figure}

The launch pad surface with post-mortem particle launch sites is shown in Fig. \ref{fig:VvsE}, where the variation of the extent of permanent expansion/deformation depends on the laser pulse energy. Above certain pulse energies, depending on the expansion-layer material and thickness (400 mJ for 40 µm Cu and 625 mJ for 100 µm Al), the foils fail and rupture around the ablation area, becoming flyers that impact the target along with the particle. When accelerating particles with larger inertias (e.g., 80 µm tungsten carbide and 100 µm stainless steel), the metal foils have been found to puncture at the particle contact point during expansion. In the current design and implementation, these are the limiting factors in achieving higher impact velocities without contaminating the target. These limitations can potentially be overcome by implementing techniques that prevent the flyer and the ablation products from reaching the target. Overall, the proposed design of the launch pad assembly offers the capability of launching larger (and denser) microparticles at supersonic velocities and testing of materials at ultra-high temperatures, while being simple to fabricate and maintaining the desirable cost-effectiveness and the high-throughput nature of LIPIT experiments. Furthermore, laser-driven flyer experiments can be easily accommodated in the system by introducing a higher quality diffractive optical element and increased pulse energies. A single 2" $\times$ 2" square launch pad can accommodate up to 25 experiments.

\subsubsection{Target Mount and Resistive Heating System}

Strict optical constraints for alignment and diagnostics, as well as the microprecision requirement of the experiment dictate the way the target sample is mounted. The addition of resistive heating introduces additional restrictions to the design of the target mounts. In this setup, a cuboid target sample with dimensions of 5 mm in width, 3 mm in height, and a thickness ranging from 1.5 to 2 mm is held using a pair of 1.27 mm (0.05 in.) diameter cylindrical tungsten rods. The ends of the rods holding the target sample are wire electric discharge machined (EDM) to have semicircular cross sections for 10 mm along the length, where the flat surfaces are in contact with the sides of the target sample as shown in Fig. \ref{fig:targetMount} (a). The tungsten rods are 76.2 mm (3 in.) in length and are glued into the slots of custom designed parts that in turn are held in the slots of an XY stage (Thorlabs XYF-1). The distance between the slots can be varied, allowing the assembly to hold target samples of different sizes. The XY stage is mounted on a precision rotary stage (Thorlabs PR005), which is connected to a rotation mount (Newport RSP-1T). Together, these enable precise vertical and lateral angle adjustments, ensuring the normal impact of the projectile. The entire assembly is mounted on the optical table via linear XYZ translation stages for fine adjustments in the positioning of the target.

The resistive heating of the target sample to ultra-high temperatures is achieved using a programmable auto-range DC power supply (ITech IT6502D) with an output capacity up to 80 V and 60 A. The power supply terminals are connected to the two tungsten rods (as shown in  Fig. \ref{fig:targetMount} (a)) that hold the target and pass current that leads to its heating. The slenderness of the electrodes allows for the thermal expansion of the target and maintains contact up to temperatures close to the thermal softening of tungsten. This implementation of direct resistive heating requires the target material to be of conductive nature and have a prismatic form with dimensions on the order of millimeters for achieving proper mounting and electrode contact. Furthermore, potential chemical reactivity of the target material with the tungsten electrodes at elevated temperatures will need to be considered on a case-by-case basis.

\begin{table}
\caption{\label{tab:table1} IR camera specifications and parameters.}
\begin{ruledtabular}
\begin{tabular}{lll}
Camera/parameters&\mbox{Optris Xi400}&\mbox{Optris PI05M}\\
\hline
Field of view& 76 $\times$ 57.3 mm$^2$ & 22.1 $\times$ 29.3 mm$^2$\\
Framing rate&  80 Hz & 27 Hz \\
Optical resolution& 382 $\times$ 288 px & 382 $\times$ 288 px\\
Spectral range& 8 - 14 µm &  500 - 540 nm\\
Temperature range& -20 - 1000 °C & 900 - 2450 °C \\
Lens focal length & 20 mm & 25 mm \\
Minimum distance to target & 350 mm & 100 mm \\
IFOV& 80 µm & 100 µm \\
Accuracy& $\pm 2\%$ & $\pm 1\%$ \\

\end{tabular}
\end{ruledtabular}
\end{table}

Temperature monitoring is performed through infrared (IR) imaging of the target specimen. For experiments up to 900 °C, an Optris Xi400 (range: -20 °C - 1000 °C) IR camera is implemented, and an Optris PI05M (range: 900 °C - 2450 °C) is used for temperatures beyond that. The IR camera parameters and specifications are provided in Table \ref{tab:table1}. Due to the optical constraints intrinsic to the experiment, where the direct view to the front surface of the target is obstructed by the launch pad and the rear surface is blocked by the alignment microscope objective, the temperature measurement is done on either the top or side surfaces of the target, depending on the experiment configuration.

To establish the validity of the heating and the IR thermography methods, a calibration study is performed using POCO graphite specimens. Here, the Xi400 and PI05M cameras are positioned so as to measure the front and lateral (in this case top) surface temperatures, respectively, of the graphite specimen mounted on the target assembly (Fig. \ref{fig:tempCalib} (a), (c)). As shown in Fig. \ref{fig:tempCalib} (d), for 2.5 A and 5 A current loads, the temperatures remain below 900 °C. For the 10 A and 30 A load cases, the temperatures exceed the 900 °C lower limit of PI05M, and for these two cases, the temperature-time plots of the two measurements are time-matched to confirm their agreement. The overlap of the two temperature curves confirms that the temperatures measured by the two IR cameras agree and that measurement on a lateral surface of the target is a valid method of temperature measurement for LIPIT experiments. The variation in temperature along the width of the sample is evaluated by taking measurements along the centerline of the top surface, as shown in Fig. \ref{fig:tempCalib} (a). The temperature profile is found to reach a minimum midway between the electrodes (Fig. \ref{fig:tempCalib} (b)). A slight bias in the temperature distribution between two contact points is observed, which is believed to be due to the difference in the two electrode-target contact pressures. However, both of these variations fall off as the target reaches the maximum temperature for the given current, with the highest temperature difference between any two points being less than $2\%$ at the time of the experiment. This variation is acceptable for microparticle impact experiments, where it is negligible within the length scale and in the vicinity of the impact zone. It is also noted that, as seen from the frontal camera measurement (Fig. \ref{fig:tempCalib} c), the temperature of the tungsten electrodes rapidly decreases away from the contact area. It has further been demonstrated in a separate study that the current method enables high-temperature experiments at varying obliquities (impact angles) \cite{ochilov2026coupled}. To further support and inform the heating and temperature measurement methods, relevant COMSOL Multiphysics® simulations are performed and are discussed in Appendix A.

\begin{figure*}
\includegraphics[width=0.85\textwidth]{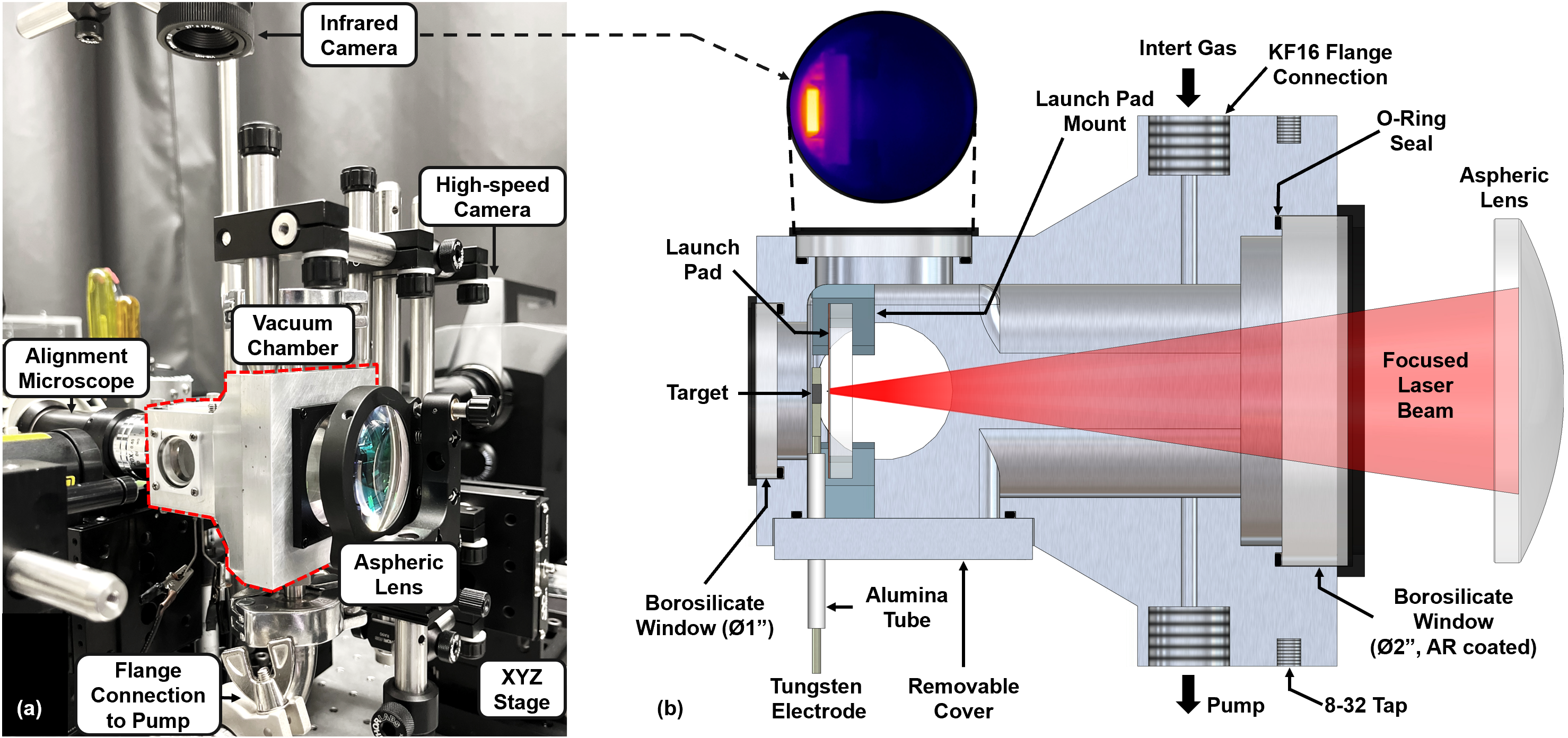}
	\caption{Vacuum chamber: a) integrated into the experiment setup; b) Illustration of the symmetry midplane cross section.}
	\label{fig:chamber}
\end{figure*}

\subsubsection{Vacuum Chamber: Design and Integration}

The formation of an oxide layer on material surfaces at elevated temperatures can drastically alter their properties. Therefore, the capability to decouple this effect from material behavior by eliminating oxidation is critical when investigating the high-temperature mechanical response of materials. This is typically achieved by conducting experiments in an inert gas environment or under vacuum. In this study, a vacuum chamber that facilitates experiments in both inert gas and high-vacuum environments is developed and integrated into the experiment setup (Fig. \ref{fig:chamber}(a)). The chamber is designed to meet strict optical constraints and requirements for microscale alignment and diagnostics while accommodating the repeated heating of the target sample to ultra-high temperatures under high vacuum. To achieve this, the body of the chamber, machined from a single block of Aluminum 6061, is designed to comprise a cubic and a divergent section. Figure \ref{fig:chamber} shows the symmetry plane cross-section of the vacuum chamber (with launch pad and target loaded), and its positioning relative to the aspheric lens. The cubic section houses the launch pad and target assemblies, and where the side, top, and rear walls are equipped with 25.4 mm (1") diameter optical windows for high-speed imaging, IR imaging, and particle alignment, respectively (Fig. \ref{fig:chamber}(b)). For IR imaging, the choice of window material requires special consideration, as it needs to have high transmissivity at the spectral range of the IR camera employed.  In this study, the window used is either a 3.175 mm thick AR-coated borosilicate glass (for PI05M) or a 5 mm thick Potassium Bromide (Thorlabs WG10255) (for Xi400), depending on the desired temperature in experiments. The transmissivity of the window material (0.9 for both the borosilicate and Potassium Bromide within their respective applications wavelengths) is accounted for in IR thermography measurements. The bottom wall of the cubic section is replaced with a removable cover that holds the tungsten electrodes and provides access to the interior. The internal dimensions of the cubic section are 30 $\times$ 30 $\times$ 28 mm. The divergent section of the chamber (Fig. \ref{fig:chamber}(c)) accommodates two KF16 flange connections, one to the vacuum pump and the other for the optional inert gas supply. Filling of the chamber with inert gas is expected to induce drag effects similar to that of experiments outside of the chamber (in the atmosphere), as discussed in \cite{veysset2020laser}. The convective heating of the launchpad is expected to be minimal given the short heating periods and the capability of variable launch pad positioning distance. The front of the divergent section holds the larger 50.8 mm (2") diameter, 9.525 mm (2/8") thick high-transmissivity, anti-reflective coated circular borosilicate glass through which the focused driving laser pulse enters. All windows and the removable cover are sealed using O-rings.
Here, a 25.4 mm (1") diameter, 3.175 mm (1/8") thick borosilicate glass is used as the launch pad substrate, and the complete launch pad assembly is mounted inside the chamber using 3D-printed gripper arms. The four gripper arms (two of which are fixed on the removable cover) extend from each edge of the interior of the chamber and are designed not to interfere with the lighting and imaging of the experiment. The positioning of the launch pad along the direction of the driving laser beam can be varied and requires consideration of the working distances of the rear alignment and the high-speed imaging microscope objectives. This is done by varying the position of the gripper arms. In-plane variation of the relative positioning of the target and launch pad can be achieved by moving the target along the length of the electrodes and by rotating the launch pad in place. A single launch pad with a diameter of 25.4 mm (1") can accommodate up to 5 experiments.

\begin{figure*}
\includegraphics[width=1\textwidth]{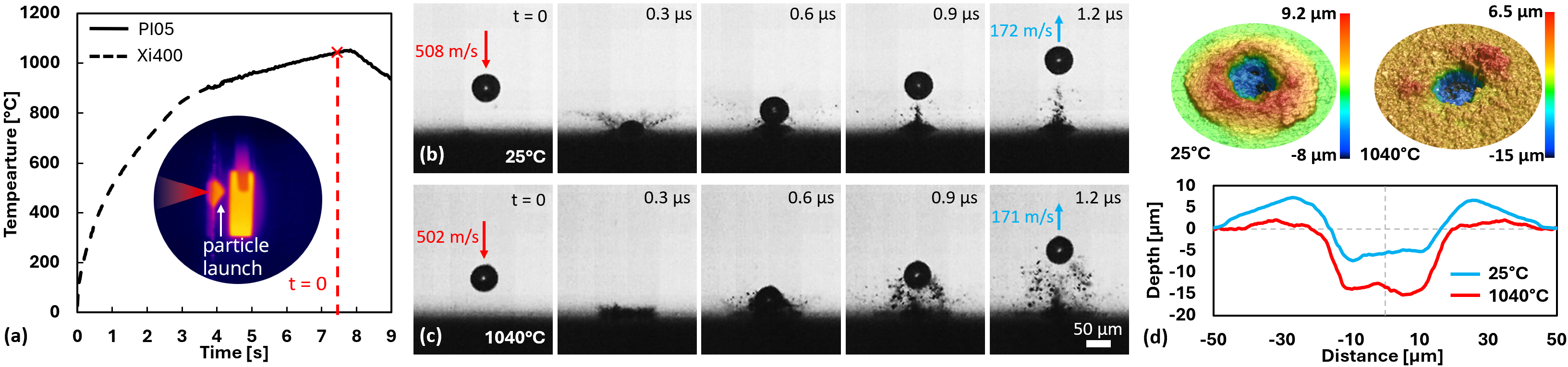}
	\caption{Experiments in atmosphere. (a) Temperature vs. time plot for the  1040 °C experiment. (b) $\&$ (c) 300 ns interframe images for the room and high-temperature experiments, respectively. (d) Laser confocal microscopy of impact sites using Keyence VK-X3000.}
	\label{fig:RoomVs1000C}
\end{figure*}

In an identical manner to the target mount assembly described in Section 3, the target here is held between the two tungsten electrodes of the same design. To both electrically and thermally insulate the rods from the rest of the chamber, they are fed through 25 mm long cylindrical alumina tubes, which are in turn fed through the removable cover. The distance between the two feedthrough holes is predetermined based on the target sample width of 5 mm. The contact surfaces between the rods, alumina tubes, and the removable cover are sealed using a high-temperaute adhesive (Permabond 820). The vacuum chamber assembly is mounted onto the optical table via an XYZ micrometer stage triad for precise and independent positioning. The chamber is pumped down using a Pfeiffer HiCube Eco vacuum pump, and the minimum pressure achieved is 5$\times10^{-4}$ mbar (0.375 mTorr).

\subsubsection{High-speed Microscopic Imaging System}

In these experiments, the impact and rebound velocities are measured using high-speed microscopic imaging. A high-speed camera (Shimadzu HPV-X2) equipped with a modular tube microscope lens (Navitar) and a long-working-distance objective (Mitutoyo 10×, NA 0.28, WD 33.5 mm) is used for imaging. The camera is capable of recording images up to 10 million frames per second with a resolution of 400 $\times$ 250 pixels. Illumination is provided by a pulsed laser (Cavilux UHS Smart), which delivers up to 400 W of low-coherence light at a wavelength of 640 nm. The laser produces illumination pulses as short as 10 ns with a minimum interval of 100 ns, making it ideally suited for synchronization with the high-speed camera selected in this study. The short pulse duration minimizes motion blur during imaging of the high-speed impact events.  It is noted that the magnification used in this study enables us to obtain pixel resolutions of 2.1 µm/pixel, which helps resolve features and particles as small as 6 µm in size. 

% Therefore, the particles of sizes varying from 6 µm to above 100 µm can be used in the experiments, expanding the lower strain rate regimes accessible in LIPIT experiments.

\subsection{Application: Graphite Case Study}

\subsubsection{Experiments in Atmospheric Conditions}

The microparticle impact experiments are conducted on ultra-fine grain graphite material (POCO ZXF-5Q) at room temperature (RT) and at 1040 °C under atmospheric (in air) conditions. Target mount and launch pad assemblies described in Section A3 are implemented for these experiments. The front (impact) surface of each graphite target is mirror polished (up to 2500 grit) before being mounted. Nominally 60 µm diameter alumina microsphere projectiles are accelerated to above 500 m/s before impacting the target sample. Each experiment is recorded at 10 million FPS (100 ns interframe time), and the 300 ns increment snapshots of the impact evolution are presented in Fig. \ref{fig:RoomVs1000C} (b) and (c). The alumina projectiles remain intact owing to their high strength and rebound with a residual velocity. The surface temperature-time graph of the resistively heated target specimen is shown in Fig. \ref{fig:RoomVs1000C} (a), where it can be seen that the target temperature of 1040 °C is reached within 7.4 seconds of heating. Although this is a relatively short heating time during which no macroscopically significant oxidation weight loss occurs \cite{segletes1973oxidation, balden2005oxidative}, the surface effect at the microscale is substantial. This is evident in Fig. \ref{fig:RoomVs1000C} (c), where the laser confocal microscopy (Keyence VK-X3000) of impact sites (Fig. \ref{fig:RoomVs1000C} d) shows striking differences in both the surface roughness and crater morphologies of the two target specimens. The heated target exhibits increased surface roughness due to oxidation and has a maximum crater penetration depth of 15.2 µm (over double the value for the room temperature experiment, where it is 7.3 µm), as well as significantly decreased piling at the crater rim compared to the room temperature impact crater profile. It is also noted that such short heating times ensure the particle does not reach temperatures close to that of the target before impact, and the target cooling at impact is considered negligible, with contact times being < 500 ns. If experiments require the equilibration of target and particle temperatures, it is necessary that they remain lower than the melting point of the metal foil used as the driving layer.

\begin{figure*}
\includegraphics[width=0.97\textwidth]{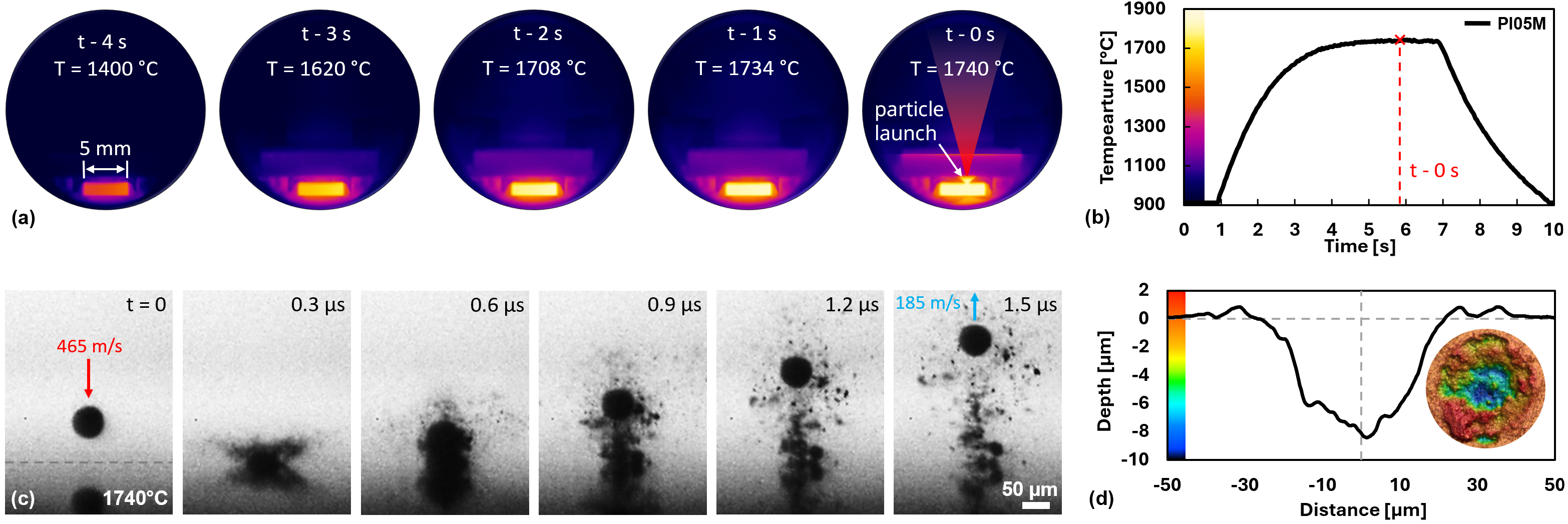}
	\caption{Ultra-high temperature experiment under high vacuum at 1740 °C. (a) IR thermography snapshots of the heating of the target. (b) Temperature vs. time plot. (c) 300 ns increment high-speed imaging frames - 60 µm particle launched by a 100 µm aluminum foil driving layer at 450 mJ pulse.} (d)  Laser confocal microscopy of impact site using Keyence VK-X3000.
	\label{fig:pghv}
\end{figure*}

Higher-temperature graphite experiments were conducted in air, but post-mortem scanning and crater diagnostics were infeasible because the graphite target experienced excessive oxidation/nitridation and pitting above 1300 °C. This demonstrates the necessity of experiments in an inert/vacuum environment, where ultra-high-temperature testing can be conducted while isolating the effects of temperature from oxidation when assessing the cratering and erosion behavior of materials.

\subsubsection{Experiments under Vacuum}

For these experiments, the vacuum chamber described in Section A4 is integrated into the setup. Once the launch pad (with particles placed on the surface) is mounted in the chamber, particle alignment is done by translating the entire chamber using the XYZ stage while keeping the launch pad at the correct distance ($\approx$ 100 mm) away from the focusing lens to maintain a consistent laser spot size. After the particle is aligned with the laser ablation spot (launch site), the removable cover with the target mounted between the tungsten electrodes is inserted in place, the cover is secured with screws, and the chamber is sealed. Next, the illumination laser, along with the high-speed and IR cameras, is aligned before the vacuum is pulled. Once the desired pressure is reached, the target is heated, and a pulse is released from the driving laser, which in turn triggers the high-speed imaging process.

Figure \ref{fig:pghv} (a) shows one-second increment IR thermography snapshots leading up to the experiment, where the temperature is measured on the top surface of the target sample and plotted against time in Fig. \ref{fig:pghv} (b). For this experiment, the target temperature of 1740 °C is reached in less than 6 seconds under 25 A current loading. Figure \ref{fig:pghv} (c) shows the 300 ns increment high-speed imaging frames of the experiment, where the 60 µm alumina microsphere impacts the graphite target at 465 m/s and rebounds at 185 m/s. Laser confocal microscopy of the impact surface shown in Fig. \ref{fig:pghv} (d) reveals a close to pristine and polished finish, indicating the absence of oxidation/pitting of graphite. The crater morphology exhibits a unique form that is not observed in room temperature and atmospheric high-temperature experiments.
Although in-depth characterization of graphite erosion at extreme temperatures will require an extensive number of experiments and further detailed investigation, the experiments discussed above demonstrate the capability of the developed setup in accomplishing such a study.

Target temperatures exceeding 2000 °C are achieved under vacuum by increasing the current load to above 40 A. The launch pad assembly, when positioned an adequate distance ($\ge$ 5 mm) away from the target, is able to accelerate particles to supersonic velocities at these experiment temperatures. However, the combination of significant thermal expansion of the target and the softening of tungsten electrodes results in the deformation of the electrodes near the contact surfaces. This leads to the misalignment of the target and the eventual loss of electrode contact.

\section{Summary and Conclusion}\label{Conclusions}
In this study, a LIPIT system has been developed that is capable of high-speed microparticle impact experiments at ultra-high temperatures in air and under vacuum. This is achieved through the development of a launch pad design utilizing an aluminum/copper foil-based driving layer (with thicknesses ranging from 35 µm to 100 µm) and the integration of a resistive heating system. The proposed launch pad design has been shown to accelerate particles of sizes on the order of 100 µm to supersonic velocities while withstanding heating of the target material to 1740 °C. This launch pad design is both simple and easy to fabricate, providing distinct benefits that enhance the inherent high-throughput characteristics and accessibility of LIPIT experiments while facilitating experiments at extreme temperatures.

Methods for resistive heating using tungsten electrodes and the infrared thermography of the target are introduced as streamlined and robust procedures. Their reliability is verified through calibration experiments and numerical simulations of graphite heating. The maximum achievable temperature limit in the setup is determined by the thermal softening and deformation of the tungsten electrodes due to the thermal expansion of the target at 2000°C.

An optically accessible vacuum chamber capable of high-vacuum and inert-gas environments has been developed and integrated into the setup as a modular system. The chamber is shown to accommodate experiments at ultrahigh temperatures under vacuum pressures as low as 5$\times10^{-4}$ mbar (0.375 mTorr).

These contributions expand the operational envelope of LIPIT systems, increasing access to particle velocity regimes across size and density scales, extreme target temperatures, and inert or high-vacuum experiment environments.
\begin{acknowledgements}

This research was conducted with support from the grant N000142512178 from the Office of Naval Research, Defense University Research Instrumentation Program (DURIP) grant N00014-23-1-2697, and FY2020 MURI grant N00014-20-1-2682. We want to thank Prof. Thomas E. Schwartzentruber at the University of Minnesota for his valuable guidance and the many motivating discussions that helped shape this project. Portions of this work were conducted at the Minnesota Nano Center (which is supported by the National Science Foundation through the National Nanotechnology Coordinated Infrastructure (NNCI) under Award Number ECCS-2025124) and with the help of the University of Minnesota CSE Machine Shop.

\end{acknowledgements}

\section*{Data Availability Statement}

The data that support the findings of this study are available from the corresponding author upon reasonable request.

\newpage

\appendix

\begin{table}
\caption{\label{tab:table2} Graphite material properties.}
\begin{ruledtabular}
\begin{tabular}{ll}
Property&\mbox{POCO Graphite ZXF-5Q}\\
\hline
Density & 1.78 g/cm$^3$ (0.064 lb/in$^3$)\\
Porosity &  20$\%$ volume\\
Compressive strength & 175 MPa (23,600 psi)\\
Electrical resistivity & 1950 $\mu\Omega-cm$ (770 $\mu\Omega-in$)\\
Thermal conductivity & 70 W/m-K (40 Btu-ft/hr/ft$^2$°F)\\
Oxidation threshold & 450°C (840°F)\\
Particle size& 1 µm (40 µin)\\
Pore size & 0.3 µm (12 µin) \\

\end{tabular}
\end{ruledtabular}
\end{table}

\section{COMSOL Simulations of Resistive Heating}\label{COMSOL}

 \begin{figure*}
\includegraphics[width=0.8\textwidth]{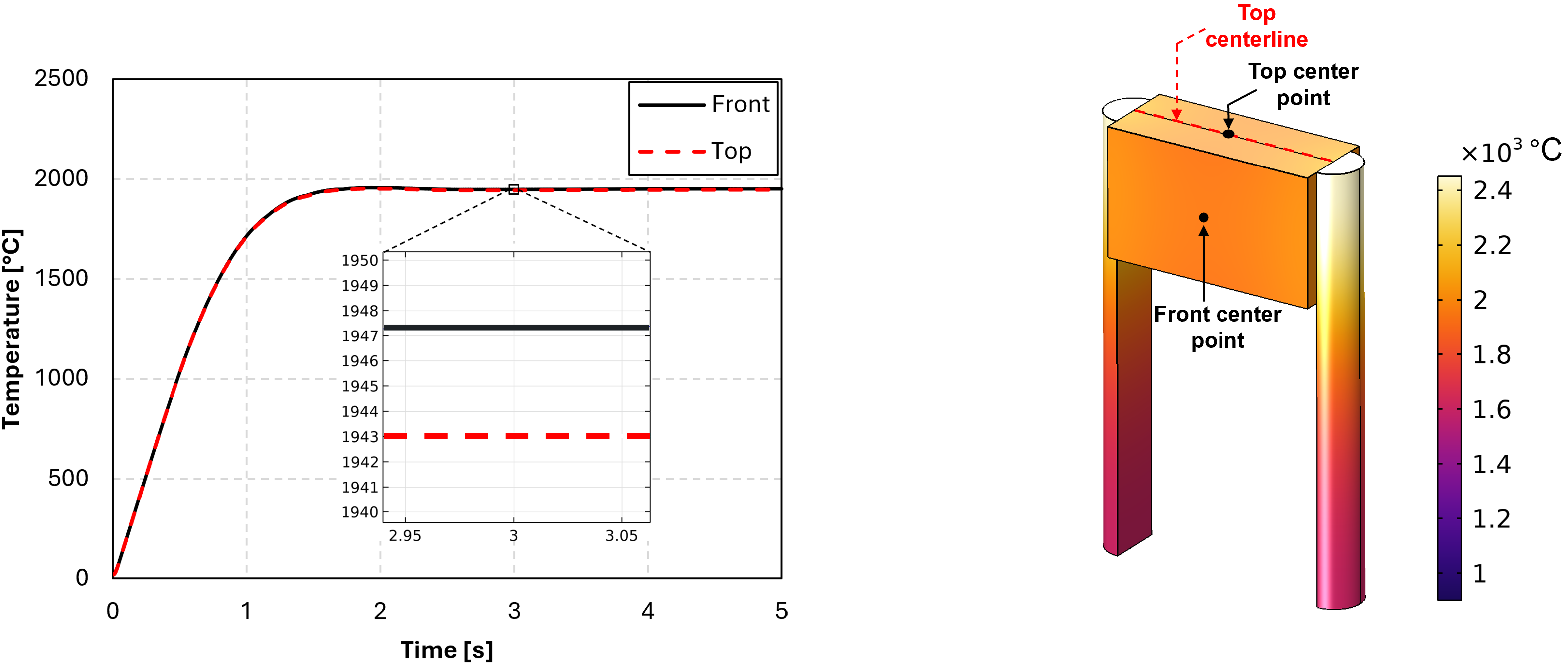}
	\caption{COMSOL Multiphysics® simulation temperature vs time history comparison of the front and top center points and the isometric view of the simulation temperature distribution at 3 s.}
	\label{fig:comsol-iso}
\end{figure*}

\begin{figure}
\includegraphics[width=0.43\textwidth]{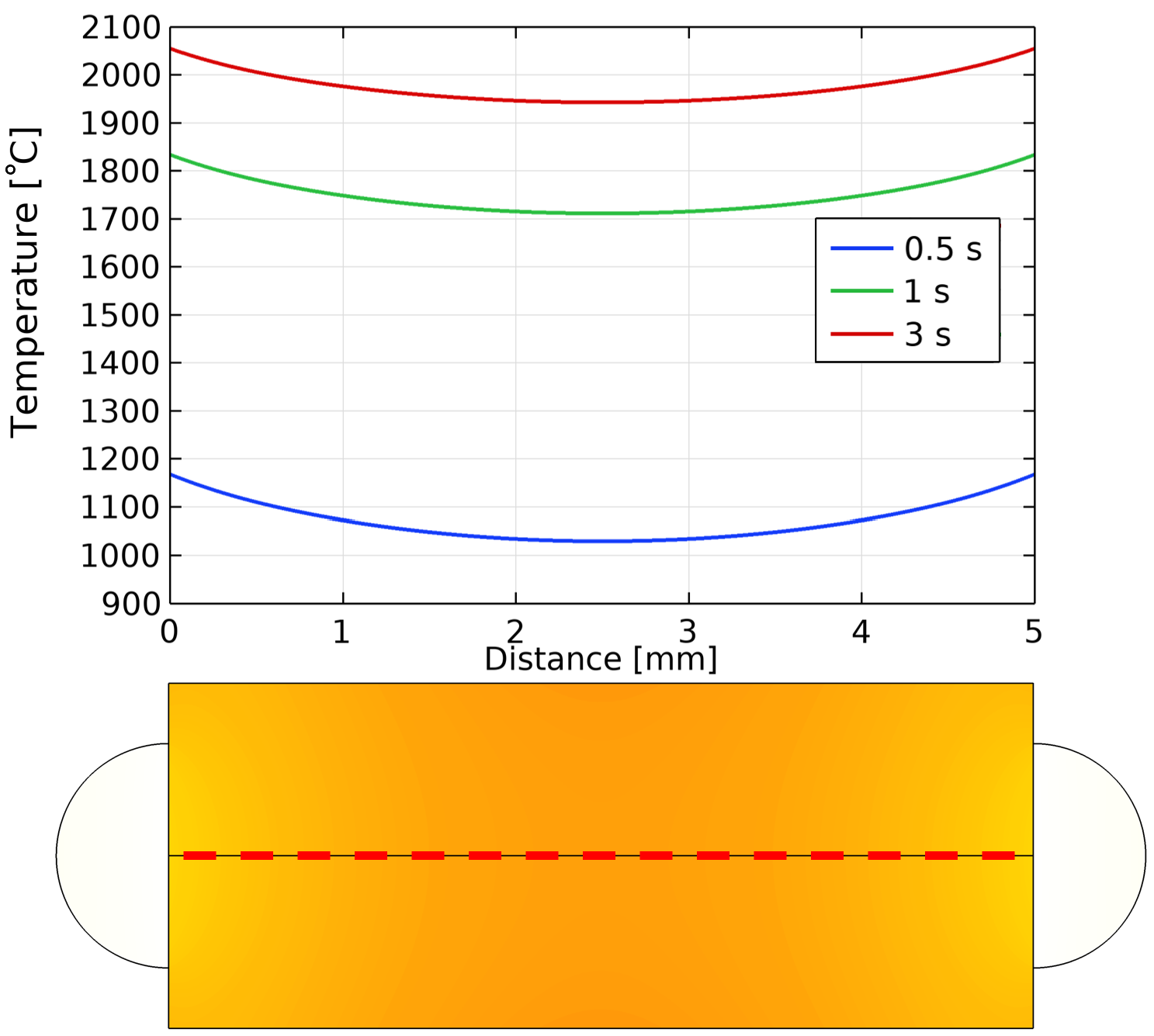}
	\caption{COMSOL Multiphysics® simulation variation of temperature along the top surface centerline at 0.5, 1, and 3 s of heating.}
	\label{fig:comsol-lin}
\end{figure}

COMSOL Multiphysics® was used to inform and validate the target heating and temperature measurement methods implemented in the experiments. The target (POCO ZXF-5Q Graphite) and a 10 mm section of the tungsten rods were modeled with resistive heating at a fixed current of 10 A. Conduction and radiation were simulated, while convection was seen to be negligible. The material properties used (provided in Table \ref{tab:table2}) were those at room temperature and were assumed to be invariant with temperature. This approximation is adopted as it is not the exact temperature values that are of interest but rather the impact of experiment geometry and resistive heating on the temperature gradients that form on the target. The simulation showed gradients similar to those observed during the experiment, where the center of the target was the coldest point with hot spots near the interface with the tungsten rods (Fig. \ref{fig:comsol-lin}). Furthermore, it is also found that the temperature difference between the center points on the top and front surfaces is less than 5 °C once the target has reached maximum temperature approaching 2000 °C as shown in Fig. \ref{fig:comsol-iso}.

\newpage

\bibliographystyle{apsrev4-1}
\bibliography{References.bib}

\end{document}